\documentstyle[epsfig,aps,prd]{revtex}
\begin{document}
\bibliographystyle{plain}
\thispagestyle{empty}
\begin{flushright}
IISc-CTS-24/00\\
hep-ph/0011274
\end{flushright}
\bigskip
\bigskip
\bigskip
\begin{center}
\large{\bf   
Infrared Fixed Point Structure in Minimal Supersymmetric Standard Model
with Baryon and Lepton Number Violation}

\vskip 2.5cm

B. Ananthanarayan\\
Centre for Theoretical Studies,\\
Indian Institute of Science, Bangalore 560 012, India 

\vskip 1.0cm

P. N. Pandita\\
Department of Physics,\\
North-Eastern Hill University, Shillong 793 022, India

\end{center}

\vskip 4.0cm

\begin{abstract}
We study in detail the renomalization group evolution of Yukawa couplings
and soft supersymmetry breaking trilinear couplings in the minimal
supersymmetric standard model with baryon and lepton number violation.
We obtain the exact solutions of these equations in a closed form, and 
then depict the infrared fixed point structure of the third generation
Yukawa couplings and the highest generation baryon and lepton number
violating couplings.  Approximate analytical solutions for these 
Yukawa couplings and baryon and lepton number violating couplings,
and the soft supersymmetry breaking couplings  
are obtained in terms of  their initial values at the unification scale.  
We then numerically study the infrared fixed surfaces of the model,
and illustrate the approach to the fixed points.
\end{abstract}

\bigskip

\vskip 1.0cm

PACS number(s):  11.10.Hi, 11.30.Fs, 12.60.Jv


\vskip 1.0cm

\begin{flushright}
Typset using REVTeX
\end{flushright}

\newpage

\section{INTRODUCTION}

Considerable attention has been focussed on the infrared fixed point
behavior~\cite{bsmw1} of the standard model (SM) and its extensions, 
especially the minimal supersymmetric standard model (MSSM). 
This is because there may be
a stage of unification beyond the SM, and if so, it then becomes important
to perform the radiative corrections in determining all the dimension
$\le 4$ terms in the lagrangian. This can be achieved by using the 
renormalization group equations to find the values of parameters
at the electroweak scale, given their values at the unification
scale. As such, much effort has been devoted to the study of  
the evolution of  various dimensionless Yukawa couplings
in the SM, and its minimal supersymmetric extension, the MSSM.
Using the renormalization group evolution,
one can relate the  Yukawa couplings to the gauge couplings
via the Pendleton-Ross infrared stable fixed point (IRSFP) for the top-quark
Yukawa coupling~\cite{bpggr1,mlggr1}, or via the quasi-fixed-point 
behavior~\cite{cth1}. The predictive power of the SM and its supersymmetric
extensions may, thus, be enhanced if the renormalization group (RG)
running of parameters is dominated by infrared stable fixed points (IRSFPs).
These parameters (Yukawa couplings, ratios of Yukawa couplings to gauge 
couplings, etc.) do not attain fixed point values at the weak scale, the range 
between the grand unified theory (GUT) scale and the  weak scale being too 
small for them to closely approach the fixed point. Nevertheless, the couplings
may be determined by the quasi-fixed point behavior~\cite{cth1} where the
value of the coupling at the weak scale is independent of its value at the 
GUT scale, provided the coupling at the GUT scale is large.

In supersymmetric theories there are superpartners of the ordinary 
particles in the spectrum, due to which there are additional 
Yukawa couplings~\cite{wsy} in these models which lead to 
baryon number $(B)$  or lepton number $(L)$ violation.  Often, a 
discrete symmetry~\cite{ff} called $R$-parity
$(R_p)$ is introduced to eliminate these $B$ and $L$ violating 
Yukawa couplings. However, the assumption of $R_p$ conservation in
MSSM appears to be {\it ad hoc}, since it is not required for the
internal consistency of the model.  Considerable attention has, thus, 
recently been focussed on  the study of the renormalization group evolution
of Yukawa couplings, including the baryon and lepton number
violating couplings, of the MSSM~\cite{bbpw,bapnp1,bapnp2,add1}.  
This includes 
the study of quasi-fixed-point behavior as well as the true infrared 
fixed points of the different Yukawa couplings, and the analysis of 
their stability.  It has been shown that in the Yukawa sector of the 
minimal supersymmetric standard model with baryon and lepton number
violation there is only one infrared stable fixed point.  This  corresponds
to nontrivial fixed point for the  top- and bottom-quark Yukawa couplings
and the $B$ violating coupling $\lambda''_{233}$, and  a trivial one
for the $\tau$-Yukawa coupling and the $L$ violating coupling
$\lambda_{233}$. It was shown that all other fixed points are
either unphysical or unstable in the infrared  region.  Similarly,
fixed points were obtained for the corresponding
soft supersymmetry breaking trilinear couplings as well~\cite{bapnp2}.

The purpose of the present paper is  twofold.
Firstly, we study the renormalization group evolution 
in the minimal supersymmetric standard model with baryon and lepton
number violation in  order to  obtain the exact as well
as the  approximate
analytical solutions for the Yukawa couplings and the soft supersymmetry 
breaking couplings at the weak scale given their initial values at the 
ultraviolet (UV) or the GUT scale. Second, we study the renormalization group
flow of such a system,  and  determine the infrared fixed surfaces and
infrared fixed points toward which the RG flow is attracted.

The plan of the paper is as follows. In Sec.II 
we describe the renormalization
group equations for the minimal supersymmetric standard model with 
baryon and lepton mumber violation involving the highest generations. 
We obtain the
exact solutions for the RG equations, and describe  the infrared 
fixed points for the third generation Yukawa couplings and the 
highest generation baryon and lepton number violating couplings. Here we
also study the corresponding RG equations for the soft supersymmetry
breaking trilinear couplings, and describe their infrared fixed points.
Sec.III is devoted to the study of the approximate analytical
solutions of the RG equations for the Yukawa couplings 
and the soft supersymmetry breaking
trilinear couplings. In Sec.IV we carry out a detailed numerical study of 
the infrared attractive fixed surfaces, and present their  
two dimensional 
projections to demonstrate the existence of a strongly attractive 
fixed point.  In Sec.V  we present the summary and conclusions.

\section{RENORMALIZATION GROUP EQUATIONS AND INFRARED FIXED POINTS}
\subsection{Infrared fixed points for Yukawa couplings}
In this section we recall some of the basic features of the renormalization
group evolution in the minimal 
supersymmetric standard model with baryon and lepton number violation, and
obtain the infrared fixed points of the Yukawa couplings
and the soft supersymmetry breaking trilinear couplings of the model.
The superpotential of the model is written as
\begin{equation}\label{mssmsuperpotential}
W= (h_U)_{ab} Q^a_L \overline{U}^b_R H_2
+ (h_D)_{ab} Q^a_L \overline{D}^b_R H_1
+ (h_E)_{ab} L^a_L \overline{E}^b_R H_1 + \mu H_1 H_2,
 \end{equation}
where $L,\, Q, \, \overline{E},\, \overline{D},\, \overline{U}$
denote the lepton and quark doublets, and anti-lepon singlet, d-type
anti-quark singlet and u-type anti-quark singlet,  respectively.
In Eq.~(\ref{mssmsuperpotential}), $(h_U)_{ab}$, $(h_D)_{ab}$
and $(h_E)_{ab}$ are the Yukawa coupling matrices, with $a,\, b,\, c$
as the generation indices.
Gauge invariance, supersymmetry and renormalizability allow the addition of
the following $L$ and $B$ violating terms to the
MSSM superpotential (\ref{mssmsuperpotential}):
\begin{eqnarray}
W_L &=& {1\over 2}\lambda_{abc} L^a_L L^b_L \overline{E}^c_R
+ \lambda'_{abc} L^a_L Q^b_L\overline{D}^c_R
+ \mu_i L_i H_2,  \label{Lviolating} \\
W_B &=& {1\over 2}\lambda''_{abc} \overline{D}^a_R
\overline{D}^b_R \overline{U}^c_R,   \label{Bviolating}
\end{eqnarray}
respectively. In this paper we shall consider, 
apart from the Yukawa couplings in (\ref{mssmsuperpotential}), the
dimensionless Yukawa couplings $\lambda_{abc}$,
$\lambda'_{abc}$ and $\lambda''_{abc}$ only, and ignore the 
dimensionful couplings $\mu$ and $\mu_i$.  The  couplings
$\lambda_{abc}$ and $\lambda''_{abc}$ are antisymmetric in their
first two indices due to $SU(2)_L$ and $SU(3)_C$ group structures,
respectively.  Corresponding to the terms in the superpotentials
(\ref{mssmsuperpotential}), (\ref{Lviolating}) and (\ref{Bviolating}),
there are  soft supersymmetry breaking trilinear terms
which can be written as
\begin{eqnarray}
-V_{\rm{soft}} &=&
\left[(A_U)_{ab}(h_U)_{ab} \tilde{Q}^a_L \tilde{\overline{U}}^b_R H_2
+ (A_D)_{ab}(h_D)_{ab} \tilde{Q}^a_L \tilde{\overline{D}}^b_R H_1
\right. \nonumber \\
&+&\left. (A_E)_{ab}(h_E)_{ab} \tilde{L}^a_L \tilde{\overline{E}}^b_R H_1\right]
\nonumber \\
& + & \left[{1\over 2}(A_\lambda)_{abc}\lambda_{abc} \tilde{L}^a_L \tilde{L}^b_L
\tilde{\overline{E}}^c_R
+ (A_{\lambda'})_{abc}\lambda'_{abc} \tilde{L}^a_L \tilde{Q}^b_L
\tilde{\overline{D}}^c_R \right]\nonumber\\ 
&+&
 \left[ {1\over 2}(A_{\lambda''})_{abc}\lambda''_{abc} \tilde{\overline{D}}^a_R
\tilde{\overline{D}}^b_R \tilde{\overline{U}}^c_R \right],  \label{soft}
\end{eqnarray}
where a tilde denotes the scalar component of the chiral superfield, and
the notation for the scalar component of the Higgs superfield is the same
as that of the corresponding superfield. In addition there are
soft supersymmetry breaking gaugino
mass terms with the masses $M_i$ with $i = 1, 2, 3,$
corresponding to the gauge groups $U(1)_Y, \, SU(2)_L, \,$ and
$SU(3)_C$, respectively.

Since the third generation Yukawa couplings are the dominant couplings in the
superpotential (\ref{mssmsuperpotential}), we shall
retain only the elements $(h_U)_{33}\equiv h_t$,
$(h_D)_{33} \equiv h_b$, $(h_L)_{33} \equiv h_\tau$
in each of the Yukawa couplings matrices $h_U,\, h_D,
\, h_L$, setting all other elements equal to zero.  Furthermore, since
there are
36 independent $L$ violating trilinear couplings $\lambda_{abc}$ and
$\lambda'_{abc}$ in (\ref{Lviolating}), and
9 independent $B$ violating couplings $\lambda''_{abc}$ in
the baryon number violating superpotential (\ref{Bviolating}),
we would have to consider 39 coupled nonlinear evolution equations
for the $L$ violating case and 12 coupled nonlinear equations for the $B$
violating case, respectively. Thus, there is a clear need 
for a radical
simplification of the evolution  equations before we can  study the
RG evolution of the Yukawa couplings in the MSSM
with  $B$ and $L$ violation. Motivated by the generational
hierarchy of the conventional Higgs couplings, we shall assume that an
analogous hierarchy amongst the different generations of $B$ and $L$
violating couplings exists.  Thus, we shall retain only the
couplings $\lambda_{233}, \, \lambda'_{333},\,  \lambda''_{233}$,
and neglect the rest~\cite{bapnp2}.  We note that $B$ and $L$
violating couplings
to higher generations evolve more strongly because of larger Higgs
couplings in their evolution equations, and hence could take larger
values than the corresponding couplings to the lighter generations.
We also note that the experimental upper limits are stronger for the
$B$ and $L$ violating couplings with lower indices~\cite{barbier}.
With these assumptions we can
write the relevant renormalization group equations as~\cite{bapnp2}
\begin{eqnarray}
16 \pi^2 {dh_t^2\over d(-\ln\, \mu^2)} &=&
h_t^2\left({16\over 3} g_3^2 + 3 g_2^2 + {13\over 15} g_1^2
- 6 h_t^2 -  h_b^2 - \lambda'^2_{333} - 2\lambda''^2_{233} \right),  
\label{htequation} \\
16 \pi^2 {dh_b^2\over d(-\ln\, \mu^2)} &=&
h_b^2\left({16\over 3} g_3^2 + 3 g_2^2 + {7\over 15} g_1^2
- h_t^2 - 6h_b^2 - h_\tau^2 - 6\lambda'^2_{333}  
- 2\lambda''^2_{233} \right),  \label{hbequation} \\
16 \pi^2 {dh_\tau^2\over d(-\ln\, \mu^2)} &=&
h_\tau^2\left(3 g_2^2 + {9\over 5} g_1^2  -3 h_b^2 - 4 h_\tau^2 
- 4\lambda^2_{233} 
- 3\lambda'^2_{333} \right),  \label{htauequation} \\
16 \pi^2 {d\lambda^2_{233}\over d(-\ln\, \mu^2)} &=&
\lambda^2_{233}\left( 3 g_2^2 + {9\over 5} g_1^2
- 4h_{\tau}^2 - 4\lambda^2_{233} - 3\lambda''^2_{233}\right), 
\label{lambdaequation}\\
16 \pi^2 {d\lambda'^2_{333}\over d(-\ln\, \mu^2)} &=&
\lambda'^2_{333}\left({16\over 3} g_3^2 + 3 g_2^2 + {7\over 15} g_1^2
-  h_t^2 - 6h_b^2 - h_{\tau}^2 - \lambda_{233}^2 - 6\lambda'^2_{333} 
- 2 \lambda''^2_{233}\right), \label{lambdaprimeequation}\\
16 \pi^2 {d\lambda''^2_{233}\over d(-\ln\, \mu^2)} &=&
\lambda''^2_{233}\left(8 g_3^2 + {4\over 5} g_1^2
- 2 h_t^2 - 2h_b^2 - 2\lambda'^2_{333} - 6\lambda''^2_{233}\right).  
\label{lambdadoubleprimeequation}
\end{eqnarray}
where $g_1, g_2, g_3$ are the gauge couplings of $U(1)_Y$ 
(in the GUT normalization),
$SU(2)_L$ and $SU(3)_C$ gauge groups, respevtively, 
and $\mu$ is the running mass scale.
We note that within the context of grand unified theories, one is led
to the situation where baryon and lepton number violating Yukawa couplings
may be related at the GUT scale, and one may no longer be able to set
one or the other arbitrarily to zero. {\it We, therefore, include both,
the baryon and the lepton number violating couplings, in our RG equations.}
The evolution equations for the gauge couplings are not affected by the
presence of $B$ and $L$
violating couplings at the one-loop level, and can be
written, in the usual notation,  as
\begin{equation} \label{gaugerge}
16\pi^2 {dg_i^2\over d(-\ln\mu^2)} = -b_i g_i^4, \, \, \, i=1, \, 2, \, 3,
\end{equation}
with
\begin{equation}
b_i = ({{33}\over{5}}, \, \, 1, \, \, -3).
\end{equation}
The corresponding one-loop renormalization group equations for the gaugino
masses $M_i,\, i=1, \, 2, \, 3$ can be written as
\begin{equation} \label{gauginorge}
16 \pi^2 {d M_i^2\over d(-\ln\mu^2)}=  - 2  b_i g_i^2 M_i^2.
\end{equation}
Defining 
\begin{equation}
\tilde Y_i =  {h_i^2\over {16{\pi}^2}}, \, \, \, 
i = t, \, \, b, \, \, \tau, \hspace{2cm} 
\tilde Y = {\lambda^2_{233}\over {16{\pi}^2}},  \label{def1}
\end{equation}
\begin{equation} 
\tilde Y' = {\lambda'^2_{333}\over {16{\pi}^2}},  \hspace{2cm}
\tilde Y'' = {\lambda''^2_{233}\over {16{\pi}^2}}, \label{def2}
\end{equation}
the solution of the RG equations (\ref{htequation}) - 
(\ref{lambdadoubleprimeequation})
for the Yukawa and the $B$ and $L$ violating
couplings can be written in a closed form~\cite{auberson1}
\begin{equation}
\tilde Y_k(t) = { {\tilde Y_k(0) F_k(t)}\over{1 + a_{kk}\tilde Y_k(0)\int_0^t
F_k(t') dt'}}, \, \, \, \, \, \, \,
t = ln({{M_X^2}\over{\mu^2}}), \label{yukawasolution}
\end{equation}
where $M_X$ is some large initial scale, and 
where $\tilde Y_k$ stands for the functions $\tilde Y_i~   
(i = t, \,  b, \,  \tau),  \,  \tilde Y, \,
\tilde Y'$, and  $\tilde Y''$.
Analogous notation holds for the functions $F_k$.
The quantities $a_{kk}$ are the 
diagonal elements of the wave function anamolous dimension matrix, 
and are given by
\begin{equation}
a_{kk} = \{6, \, 6, \, 4, \, 4, \, 6, \,  6\}, \label{aquantities}
\end{equation} 
and the functions $F_k$ are given by the set of integral equations
\begin{small}
\begin{eqnarray}
F_t(t) &=& {{E_t(t)}\over{(1 + 6 \tilde Y_b(0)\int_0^t F_b(t') dt')^{1/6} 
(1 + 6 \tilde Y'(0)\int_0^t F'(t') dt')^{1/6} 
(1 + 6 \tilde Y''(0)\int_0^t F''(t') dt')^{1/3}}}, \label{ftequation}\\
F_b(t) &=& {{E_b(t)}\over{(1 + 6 \tilde Y_t(0)\int_0^t F_t(t') dt')^{1/6} 
(1 + 4 \tilde Y_{\tau}(0)\int_0^t F_{\tau}(t') dt')^{1/4}
(1 + 6 \tilde Y'(0)\int_0^t F'(t') dt')^{1/6} 
(1 + 6 \tilde Y''(0)\int_0^t F''(t') dt')^{1/3}}}, \label{fbequation}\\
F_{\tau}(t) &=& {{E_{\tau(t)}}\over{(1 + 6 \tilde Y_b(0)\int_0^t F_b(t') dt')^{1/2} 
(1 + 6 \tilde Y(0)\int_0^t F(t') dt')
(1 + 6 \tilde Y'(0)\int_0^t F'(t') dt')^{1/2}}}, \label{ftauequation}\\
F(t) &=& {{E(t)}\over{
(1 + 4 \tilde Y_{\tau}(0)\int_0^t F_{\tau}(t') dt')
(1 + 6 \tilde Y''(0)\int_0^t F''(t') dt')^{1/2}}}, \label{fequation}\\
F'(t) &=& {{E'(t)}\over{(1 + 6 \tilde Y_t(0)\int_0^t F_t(t') dt')^{1/6}
(1 + 6 \tilde Y_b(0)\int_0^t F_t(t') dt')
(1 + 4 \tilde Y_{\tau}(0)\int_0^t F_{\tau}(t') dt')^{1/4}
(1 + 4 \tilde Y(0)\int_0^t F(t') dt')^{1/4}}} \nonumber \\
& \times & 
{{1}\over{(1 + 6 \tilde Y''(0)\int_0^t F''(t') dt')^{1/3}}}, \label{fprimeequation}\\
F''(t) &=& {{E''(t)}\over{(1 + 6 \tilde Y_t(0)\int_0^t F_t(t') dt')^{1/3}
(1 + 6 \tilde Y_b(0)\int_0^t F_t(t') dt')^{1/3}
(1 + 6 \tilde Y'(0)\int_0^t F'(t') dt')^{1/3}}}, \label{fdoubleprimeequation}
\end{eqnarray}
\end{small}
where the functions $E_k(t)$ 
$( = E_t(t), E_b(t), E_{\tau}(t), E(t),  E'(t)$ and
$E''(t) )$ are given by
\begin{equation}
E_k(t) = \prod_{i =1}^3 \left( 1 + b_i \tilde \alpha_i(0) t\right)^{c_{ki}/b_i},
\label{eequation}
\end{equation}
with
\begin{equation}
\tilde\alpha_i(0) = {{g_i^2(0)}\over{16 \pi^2}}, \, \, \, \, i = 1, \, 2, \, 3, 
\label{tildealpha}
\end{equation}
\begin{equation}
c_{ti} = \left( {{13}\over {15}}, \, 3, \, {{16}\over{ 3}} \right), \, \, \, \, \,
c_{bi} = \left( {{7}\over {15}}, \, 3, \, {{16}\over {3}} \right), \, \, \, \, \,
c_{{\tau} i} = \left( {{9}\over{ 5}}, \, 3, \, 0 \right), \, \, \, \, \,
\label{cfunctions1}
\end{equation}
\begin{equation} 
c_{{\lambda_{233}}i} = \left( {{9}\over{5}}, \, 3, \, 0 \right), 
\, \, \, \, \, \,
c_{{\lambda'_{333}}i} = \left( {{7}\over {15}}, \, 3, \, {{16}\over {3}} \right), 
\, \, \, \, \,
c_{{\lambda''_{233}}i} = \left( {{4}\over{5}}, \, 0, \, 8 \right).
\label{cfunctions2}
\end{equation}
The solutions for the RG equations (\ref{gaugerge}) for the gauge couplings 
and the 
gaugino masses (\ref{gauginorge}) are well known and will not be repeated here.
We note that (\ref{yukawasolution}) gives the exact solution for the Yukawa couplings,
while $F_k$'s in (\ref{ftequation}) - (\ref{fdoubleprimeequation}) should 
in principle be solved iteratively.

In order to study the infrared fixed points for the Yukawa couplings, it is
convenient to redefine
\begin{equation}\label{redefinitions}
R_t={h_t^2\over g_3^2}, \,\,\, R_b={h_b^2\over g_3^2}, \,\,\,
R_\tau={h_\tau^2\over g_3^2},  \,\,\, 
R={\lambda^2_{233}\over g_3^2}, \, \, \,
R'={\lambda'^2_{333}\over g_3^2}, \, \, \,
R''={\lambda''^2_{233}\over g_3^2}, \, \, \,
\end{equation}
and retaining only the $SU(3)_C$ gauge coupling constant $g_3$, we can
write the renormalization group equations
(\ref{htequation}) - (\ref{lambdadoubleprimeequation})  
for the Yukawa couplings as
\begin{eqnarray}
3 g_3^2{dR_t\over dg_3^2} & = &
R_t\left[{7\over 3} - 6 R_t - R_b - R' - 2 R'' \right], \label{rtequation}\\
3 g_3^2{dR_b\over dg_3^2} & = & 
R_b\left[{7\over 3} - R_t - 6 R_b - R_\tau - 6 R' - 2 R'' \right], \label{rbequation}  \\
3 g_3^2{dR_\tau\over dg_3^2} & = &
R_\tau\left[-3 -3 R_b -4 R_\tau -4 R - 3 R' \right], \label{rtauequation} \\
3 g_3^2{dR\over dg_3^2} & = & 
R\left[-3 -4 R_\tau -4 R - 3 R'  \right],  \label{requation} \\
3 g_3^2{dR'\over dg_3^2} & = &
R'\left[{7 \over 3} -  R_t - 6 R_b - R_\tau - R - 6 R' - 2 R'' \right],
\label{rprimeequation} \\  
3 g_3^2{dR''\over dg_3^2} & = &
R''\left[5 - 2 R_t -2 R_b -  2 R' - 6 R'' \right].
\label{rdoubleprimeequation}
\end{eqnarray}
There are no physically acceptable 
infrared fixed points for this set of RG equations with all the
couplings attaining  nontrivial values.
The only nontrivial infrared fixed points are obtained by neglecting
$R'$. Then, 
the true infrared fixed-points for these RG equations are~\cite{bapnp2}
\begin{eqnarray}
(R''^*,\, R_b^*,\, R_t^*) &=&\left({77\over 102},\,
{2 \over 17}, \, \, {2 \over 17}\right), \, \, \,  R_\tau^* = R^* = 0 
\, \, \, \, {\rm(stable)}, 
\label{fstable} \\
(R_b^*, \, R_t^*)&=&\left({1\over 3}, \, \, {1\over 3}\right), \, \, \, 
R_\tau^* = R^* = R''^* = 0 \, \, \, {\rm (unstable)}, 
\label{funstable1}\\
(R''^*, \, R_t^*)&=&\left({19\over 24}, \, \, {1\over 8}\right), \, \, \,
R_\tau^* = R^* = R_b^* = 0 \, \, \, {\rm (unstable)},
\label{funstabel2} \\
(R''^*, \, R_b^*)&=&\left({19\over 24}, \, \, {1\over 8}\right), \,
R_\tau^* = R^* = R_t^* = 0 \, \, \, {\rm (unstable)}.
\label{funstable3} 
\end{eqnarray}
We note that the $\tau$ Yukawa coupling $R_\tau$ and the lepton
number violating coupling $R$ approach trivial fixed point
values since there is no contribution from the
$SU(3)$ coupling $g_3$  in their renormalization group equations.
{\it Also, because of this reason the $B$- and $L$- violating 
couplings do not approach 
simultaneous non-trivial fixed points.}  Furthermore, the infrared 
fixed point values of the top- and bottom-quark Yukawa couplings
for the stable fixed point (\ref{fstable}) are significantly
different from the case when $B$ and $L$ is conserved~\cite{bs1}. 
Thus, the inclusion of $B$ and $L$ violation in MSSM has the 
effect of lowering the infrared fixed point values of the 
top- and bottom-quark Yukawa couplings. These are the important
conclusions of our analysis. From (\ref{fstable})
we note that the fixed point value for the 
top-quark Yukawa coupling translates into a top-quark (pole) mass 
of about $m_t \simeq 70 \sin\beta$ GeV, which is incompatible
with the measured value of~\cite{groom} the top mass, $m_t \simeq 
174$ GeV, for any value of $\tan\beta$. Thus, the stable
infrared fixed point (\ref{fstable}) is not actually realised in 
nature.

The infrared fixed points(IRFP's)
that we have discussed above are the true IRFP's
of the renormalization equations for the 
Yukawa and baryon and lepton number violating couplings. 
However, these fixed points may
not be reached in practice, the range between the large (GUT) scale
and the weak scale being too small for the ratios to approach
the fixed point values.  In that  case, the various Yukawa
couplings may be determined by quasi-fixed point behaviour~\cite{cth1}, 
where the values of various couplings at the weak scale are independent of
their values at the large scale, provided the Yukawa couplings
at the large scale are large. More precisely, in the regime where the
Yukawa couplings
$\tilde Y_t(0), \, \, \tilde Y_(0), \, \, \tilde Y_{\tau}(0), \, \,
\tilde Y(0), \, \, \tilde Y'(0), \, \, \tilde Y''(0) \, \, 
\rightarrow \infty$ with their ratios  fixed, it is legitimate to drop
$1$ in the denominators of the equations
(\ref{yukawasolution}) and (\ref{ftequation}) -- (\ref{fdoubleprimeequation})
so that the exact solutions for the Yukawa couplings approach the 
infrared quasi-fixed-point (IRQFP) defined by
\begin{equation}
\tilde Y_k^{QFP}(t) = {{F_k^{QFP}(t)}\over{a_{kk}\int_0^t F_k^{QFP}(t') dt'}}, 
\label{quasiyukawa}
\end{equation}
with
\begin{eqnarray}
F_t^{QFP}(t) &=& {{E_t(t)}\over{(\int_0^t F_b^{QFP}(t') dt')^{1/6}
(\int_0^t F'^{QFP}(t') dt')^{1/6} (\int_0^t F''^{QFP}(t') dt')^{1/3}}}, 
\label{qftequation}\\
F_b^{QFP}(t) &=& {{E_b(t)}\over{(\int_0^t F_t^{QFP}(t') dt')^{1/6}
(\int_0^t F_{\tau}^{QFP}(t') dt')^{1/4}
(\int_0^t F'^{QFP}(t') dt')^{1/6}
(\int_0^t F''^{QFP}(t') dt')^{1/3}}}, \label{qfbequation}\\
F_{\tau}^{QFP}(t) &=& {{E_{\tau(t)}}\over{(\int_0^t F_b^{QFP}(t') dt')^{1/2}
(\int_0^t F^{QFP}(t') dt')
(\int_0^t F'^{QFP}(t') dt')^{1/2}}}, \label{qftauequation}\\
F^{QFP}(t) &=& {{E(t)}\over{(\int_0^t F_{\tau}^{QFP}(t') dt')
(\int_0^t F''^{QFP}(t') dt')^{1/2}}}, \label{qfequation}\\
F'^{QFP}(t) &=& {{E'(t)}\over{(\int_0^t F_t^{QFP}(t') dt')^{1/6}
(\int_0^t F_t^{QFP}(t') dt')
(\int_0^t F_{\tau}^{QFP}(t') dt')^{1/4}
(\int_0^t F^{QFP}(t') dt')^{1/4}
(\int_0^t F''^{QFP}(t') dt')^{1/3}}},  \label{qfprimeequation}\\
F''^{QFP}(t) &=& {{E''(t)}\over{(\int_0^t F_t(t') dt')^{1/3}
(\int_0^t F_t(t') dt')^{1/3}
(\int_0^t F'(t') dt')^{1/3}}}.  \label{qfdoubleprimeequation}
\end{eqnarray}
We note that the result (\ref{quasiyukawa}) follows immediately 
by dropping $1$ from the denominator of (\ref{yukawasolution}),
whereas the results (\ref{qftequation}) -- (\ref{qfdoubleprimeequation})
follow from the corresponding equations (\ref{ftequation}) -- 
(\ref{fdoubleprimeequation}) by using an iterative 
procedure~\cite{auberson1,kazakov1}. 
We stress here that both the dependence on the initial conditions
for each Yukawa coupling as well as the dependence on the 
ratios of initial values of Yukawa couplings  have completely dropped
out of the runnings in Eqs.(\ref{quasiyukawa}) and (\ref{qftequation})
-- (\ref{qfdoubleprimeequation}). In other words, the 
quasi-fixed-points (\ref{quasiyukawa})  are independent of
whether the $B$ and $L$ violating couplings and the third generation
Yukawa couplings are unified or not. The fact that the ratios of the various
Yukawa couplings do not enter 
Eqs. (\ref{quasiyukawa}) -- (\ref{qfdoubleprimeequation}) implies that 
these results are valid for any $\tan\beta$ regime.

\subsection{Infrared fixed points for the trilinear soft supersymmetry 
breaking parameters}
We now consider the evolution equations for the soft supersymmetry
breaking trilinear parameters in the potential
(\ref{soft}).  For these trilinear parameters
we shall assume the same kind of generational
hierarchy  as was assumed for the
corresponding Yukawa couplings.  Thus,  we shall consider only the
highest generation trilinear coulings
$(A_U)_{33}\equiv A_t$,  $(A_D)_{33}\equiv A_b$, $(A_L)_{33}\equiv A_\tau$,
$(A_\lambda)_{233}\equiv A_\lambda$,
$(A_{\lambda'})_{333}\equiv A_{\lambda'}$,
$(A_{\lambda''})_{233}\equiv A_{\lambda''}$, setting all other elements
equal to zero.  As  there is only one IRSFP~(\ref{fstable}) 
in the MSSM with $B$ and $L$ violation, we shall consider the
IRFPs for the $A$ parameters corresponding to this case only, i.e.
for $A_t$, $A_b$, $A_\tau$, $A_\lambda$ and $A_{\lambda''}$.
Retaining only these parameters, and defining the ratios 
$\tilde{A}_i=A_i/M_3$
$(A_i=A_t, \, A_b, \, A_\tau, \, A_\lambda, \, A_{\lambda''})$, we can write
the relevant renormalization group
equations~\cite{bapnp2}
for $\tilde{A}_i$ as (neglecting the $SU(2)_L$ and $U(1)_Y$ 
gauge couplings):
\begin{eqnarray}
3 g_3^2{d \tilde{A}_t\over d g_3^2} & = &
\left[ {16\over 3} - (6 R_t +3) \tilde{A}_t-
R_b \tilde{A}_b -2 R'' \tilde{A}_{\lambda''} \right],  \label{atildet} \\
3 g_3^2{d \tilde{A}_b\over d g_3^2}& = &
\left[ {16\over 3} - R_t  \tilde{A}_t-
(6 R_b + 3) \tilde{A}_b - R_\tau \tilde{A}_\tau -
2R'' \tilde{A}_{\lambda''} \right]  \label{atildeb}, \\
3 g_3^2{d \tilde{A}_\tau\over d g_3^2}& = &
\left[ 
-3 R_b \tilde{A}_b - (4 R_\tau +\frac{R}{2} +3) 
\tilde{A}_\tau -\frac{7}{2} R \tilde{A}_{\lambda}
\right], \label{atildetau}\\
3 g_3^2{d \tilde{A}_{\lambda}\over d g_3^2}& = &
\left[ 
-\frac{7}{2} R_\tau  \tilde{A}_\tau -(\frac{R_\tau}{2}+4 R +3) 
\tilde{A}_{\lambda}\right],  
\label{atilde} \\
3 g_3^2{d \tilde{A}_{\lambda''}\over d g_3^2}& = &
\left[ 8 - 2 R_t  \tilde{A}_t- 2 R_b \tilde{A}_b 
-(6 R''+3) \tilde{A}_{\lambda''} \right].  \label{atildelambdadpr} 
\end{eqnarray}
One can obtain the exact solutions for  RG equations (\ref{atildet}) --
(\ref{atildelambdadpr}) for the
trilinear parameters $A_i$ analogous to the solutions (\ref{yukawasolution})
that we have obtained for the Yukawa couplings. The expressions for
these solutions are lengthy, and will not be
written down here. Instead we shall obtain approximate 
analytical solutions for these RG equations in Sec.III, and shall further
study them numerically in Sec.IV. Nevertheless, we can obtain the
true fixed points for the $\tilde A _i$ parameters easily. 
These are
\begin{equation}
(\tilde{A}_{\lambda''}^*, \, \tilde{A}_b^*, \,  \tilde{A}_t^*, \,
\tilde{A}_{\tau}^*) = \left(1, \, 1, \, 1, \, -{2\over 17}\right),
\, \, \, \tilde{A}_{\lambda}^*  = 0,
\label{afstable} \\
\end{equation}
It is straightforward  to show that this is the only infrared 
stable fixed point for the $\tilde A_i$ paprameters.
The stability of the fixed point (\ref{afstable}) also follows from
the general results connecting the stability of a set of $A$ parameters
to the stability of the corresponding set of Yukawa couplings~\cite{ijdrtj1}.
We further note that the fixed point values for $\tilde{A}_b$ and 
$\tilde{A}_t$  are the same as in MSSM with baryon number 
conservation~\cite{ijdrtj1}. However, the fixed point value for 
$\tilde{A}_{\tau}$ is affected by the presence of the 
$B$-violating parameter $\tilde{A}_{\lambda''}$.

\section{Analytical Solutions}
Having obtained  the infrared fixed points of the renormalization
group equations for the Yukawa couplings and the soft supersymmetry breaking
trilinear couplings, it is important to
determine the approach to these fixed points.
The rate of approach to the fixed points can be determined by solving the
RG equations in the neighbourhood of  the fixed points.  
Linearization of the RG equations (\ref{rtequation})--
(\ref{rdoubleprimeequation}) in the neighbourhood of the  stable 
infrared fixed point (\ref{fstable}) leads to the following approximate
analytical solution for the Yukawa and $B$ and $L$ violating couplings:
\begin{eqnarray}
R_t & = & \frac{2}{17}   
- 0.002 R_{\tau 0} \left( {g_3^2\over g_{3 0}^2}\right)^{-1} 
+ \left[ 0.5 (R_{t 0} - R_{b 0}) + 0.02 R_{\tau 0}\right]\left( {g_3^2\over
g_{3 0}^2}\right)^{-1.18} \nonumber \\ 
&+&\left[ 0.48 (R_{t 0} + R_{b 0}) - 0.02 R_{\tau 0}- 0.06 R''_0  - 0.07 \right]
\left( {g_3^2\over g_{3 0}^2}\right)^{-0.2}   \nonumber \\
&+&\left[ 0.02 (R_{t 0} + R_{b 0}) + 0.002 R_{\tau 0} + 0.06 R''_0  
- 0.05 \right] \left( {g_3^2\over
g_{3 0}^2}\right)^{-1.68},   \label{rteqn}\\
R_b & = &  \frac{2}{17}
+0.04 R_{\tau 0} \left( {g_3^2\over g_{3 0}^2}\right)^{-1} 
- \left[ 0.5 (R_{t 0} - R_{b 0}) + 0.02 R_{\tau 0}\right]\left( {g_3^2\over
g_{3 0}^2}\right)^{-1.18} \nonumber \\ 
&+&\left[ 0.48 (R_{t 0} + R_{b 0}) - 0.02 R_{\tau 0}- 0.06 R''_0  - 0.07 \right]
\left( {g_3^2\over g_{3 0}^2}\right)^{-0.2}   \nonumber \\
&+&\left[ 0.02 (R_{t 0} + R_{b 0}) + 0.002 R_{\tau 0}+ 0.06 R''_0  
- 0.05 \right] \left( {g_3^2\over g_{3 0}^2}\right)^{-1.68},  \label{rbeqn}  \\
R_\tau & = &  R_{\tau 0} \left( {g_3^2\over
g_{3 0}^2}\right)^{-1} ,   \\ \label{rtaueqn} 
R & = &  R_{0} \left( {g_3^2\over
g_{3 0}^2}\right)^{-0.2} ,  \\ \label{reqn} 
R'' & = & 0.75 - 0.005 R_{\tau 0}\left( {g_3^2\over g_{3 0}^2}\right)^{-1}
\nonumber \\
&+& \left[ -0.37 (R_{t 0} + R_{b 0}) + 0.02 R_{\tau 0} + 0.05 R''_0 +0.05
\right]\left( {g_3^2\over g_{3 0}^2}\right)^{-0.2}   \nonumber \\
&+& \left[ 0.37 (R_{t 0} + R_{b 0})  + 0.03 R_{\tau 0} + 0.95 R''_0- 0.8 \right]
\left( {g_3^2\over g_{3 0}^2}\right)^{-1.68},   \label{rdbpreqn}
\end{eqnarray}
where $R_{t 0}, R_{b 0}, R_{\tau 0}, R_{0}, R_{0}''$ are the corresponding
initial values at the large (UV) scale. These equations determine,
in the leading order, how fast the infrared fixed point (\ref{fstable})
is approached. The measure of infrared attraction is the size of the negative
power in $( {g_3^2/ g_{3 0}^2})$ in  various terms in the expression
for $R_i$.

Similarly, liearization of the RG equations (\ref{atildet}) -- 
(\ref{atildelambdadpr}) for the trilinear couplings about the 
stable infrared fixed point (\ref{afstable}) leads to the
following approximate solution for these couplings:
\begin{eqnarray}
\tilde{A}_t &=& 1 + 0.5 \left (\tilde{A}_{t 0} - \tilde{A}_{b 0}\right)
\left( {g_3^2\over g_{3 0}^2}\right)^{-1.2} 
+ 0.4 \left(\tilde{A}_{\lambda''0} - 1 \right)
\left( {g_3^2\over g_{3 0}^2}\right)^{-2.57 }\nonumber \\ 
&-& 0.5 \left(1.2 - (\tilde{A}_{t 0} + \tilde{A}_{b 0}) 
+ 0.8 \tilde{A}_{\lambda''0}\right)
 \left( {g_3^2\over g_{3 0}^2}\right)^{-1.21}, \label{ateqn} \\
\tilde{A}_b &=& 1 -  0.5 \left (\tilde{A}_{t 0} - \tilde{A}_{b 0}\right)
\left( {g_3^2\over g_{3 0}^2}\right)^{-1.2} 
+ 0.4 \left(\tilde{A}_{\lambda''0} - 1 \right)
\left( {g_3^2\over g_{3 0}^2}\right)^{-2.57 }\nonumber \\
&-& 0.5 \left(1.2 - (\tilde{A}_{t 0} + \tilde{A}_{b 0})  
+ 0.8 \tilde{A}_{\lambda''0}\right)
\left( {g_3^2\over g_{3 0}^2}\right)^{-1.21}, \label{abeqn} \\
\tilde{A}_\tau &=&  - {2\over 17} - 0.3 \left (\tilde{A}_{t 0} 
- \tilde{A}_{b 0}\right) \left( {g_3^2\over g_{3 0}^2}\right)^{-1.2}
+ 0.03 \left(\tilde{A}_{\lambda''0} - 1 \right)
\left( {g_3^2\over g_{3 0}^2}\right)^{-2.57 } \nonumber \\
&+& \left(0.47 + 0.03 \tilde{A}_{t 0} - 0.57 \tilde{A}_{b 0}
+ \tilde{A}_{\tau 0} + 0.19 \tilde{A}_{\lambda''0} \right)
\left( {g_3^2\over g_{3 0}^2}\right)^{-1} \nonumber \\
&-& 0.27 \left(1.2 - (\tilde{A}_{t 0} + \tilde{A}_{b 0})
+ 0.8 \tilde{A}_{\lambda''0}\right)
\left( {g_3^2\over g_{3 0}^2}\right)^{-1.21}, \label{ataueqn} \\
\tilde{A}_{\lambda}& = & \tilde{A}_{\lambda 0} \left( {g_3^2\over
g_{3 0}^2}\right)^{-1}, \label{aeqn} \\
\tilde{A}_{\lambda''} & = & 1 + \left(\tilde{A}_{\lambda'' 0} - 1 \right)
\left( {g_3^2\over g_{3 0}^2}\right)^{-2.57 } \nonumber \\
&+& 0.06 \left(1.2 - (\tilde{A}_{t 0} + \tilde{A}_{b 0})
+ 0.8 \tilde{A}_{\lambda''0}\right)
\left( {g_3^2\over g_{3 0}^2}\right)^{-1.21}, \label {adbpreqn}
\end{eqnarray}
where $\tilde{A}_{t 0},  \tilde{A}_{b 0}, \tilde{A}_{\tau 0},
\tilde{A}_{\lambda 0},  \tilde{A}_{\lambda'' 0}$ are the initial 
values of the $A$ parameters at the UV scale.

\section{Numerical Results}

Having obtained analytically 
the true fixed points
as well as the quasi-fixed points  for the various 
couplings, as well as the approximate
solutions to the corresponding RG equations, 
it is instructive to carry out a numerical study of these equations
in order to illustrate the discussion of the previous sections.  

Since the RG equations are coupled differential equations with
several fixed points, this implies that the system of equations
(\ref{rtequation}) -- (\ref{rdoubleprimeequation}) has an infrared
fixed surface, the $R_t - R_b - R''$ surface, $R_{\tau} = R =0$.
In principle, a three 
dimensional plot for the evolutions would reveal the infra-red
fixed surfaces, with the sole infrared stable fixed point 
being approached by the flow from different  directions.  However, such
a demonstration is rather cumbersome.
We shall instead concentrate on the two dimensional
projections of this three dimensional
surface to  illustrate the approach to the fixed point.

For the purposes of our numerical analysis,
we set the unification scale $M_X\simeq
10^{16}$ GeV, the unified gauge coupling $\alpha_G(M_X)=1/24.5$ and
a effective supersymmetry scale of $M_S=1$ TeV. We shall use the one-loop
RG equations in the numerical analysis, since  the difference
between the one-loop and two-loop results for the infrared
fixed points~\cite{codkaz1} is less than $10\%$.   
Reinserting the electroweak gauge couplings $g_1, g_2$ in the
RG equations  (\ref{rtequation}) -- (\ref{rdoubleprimeequation}),
the fixed points are no longer found to be exact. In order to determine the
approximate positions of the  fixed points, we shall maintain running
$g_1(\mu), g_2(\mu)$ and follow the prescription that the approximate
infrared fixed point in the $R_t - R_b - R''$ space is determined
as the unique point in the evolution path
from $\mu = M_X$ to $\mu = M_S$ 
which has the same value at $\mu = M_X$ and at
$\mu = M_S$. The results depend somewhat on the evolution path, and in 
particular on the scale $M_X$, the deviations from the values in
(\ref{fstable}) being larger, the larger the scale $M_X$. We note that
the idea of grand unification is not important for our
results. The infrared fixed point and the infrared fixed surface
are the properties of minimal supersymmetric standard model
with baryon and lepton number violation. Note that it is only in that they are
not exact but approximate do they depend on $M_X$.

Our main objective here  is to study the renormalization group
flow in the  coupling ratios  $R_t, R_b, R''$ from $M_X$ to $M_S$,
to display the fixed surfaces (projections of three dimensional surface onto
two dimensions), and to locate the infrared fixed point. 
We start with $h_t$ and $h_b$ each taking the values $(9, 1, 0.5)$ to ensure
a large range in the variation of these parameters and $\lambda''_{233}$
fixed at a reasonably large value of $1.1$  for the 
computations that we perform.  We evolve the Yukawa couplings down to 
the IR scale using the one-loop RG equations
and evaluate the coupling ratios $R_t, R_b, R''$.
The values of the  couplings so obtained
at the IR scale, viz., $M_S$, 
are then used as the input at the UV scale, $M_X$, and evolved down
to the IR scale and then iterated again.  This procedure 
is repeated until all the final
points meet in the infrared fixed point.
The results of these iterations are then projected onto the
$R_t - R_b$, $R_t - R''$ and $R_b - R''$ planes, respectively.
The results so obtained 
are presented in Figs.  1(a), 1(b) and 1(c),  respectively.  
It may be observed that
the approach to the fixed point of the ratio $R''$ is rather rapid
since at the outset this quantity was chosen close to it.  The
approach of the ratios $R_t$ and $R_b$ is significantly more dramatic
since they are required to approach the fixed point from rather
disparate values.  This clearly illustrates the strong attraction 
to the sole IRSFP.

The end result of the series of iterations that we have performed lead
to the following infrared fixed point values for the coupling ratios:
\begin{eqnarray}
R''^{QFP} & \simeq & 0.65, \nonumber \\
R^{QFP}_b & \simeq & 0.59, \nonumber \\
R^{QFP}_t & \simeq & 0.59.  \label{quasi}
\end{eqnarray}
This infrared fixed point and the fixed surfaces 
are as expected from the preceeding
analytical considerations with,  however, shifted position
of the fixed point due to the effect of including the
electroweak gauge couplings in our numerical iterations. It is,
in fact, the quasi fixed point for the respective coupling ratios.
The fixed point (\ref{quasi}) implies the approximate 
top-bottom Yukawa unification at all scales $\mu$.

The fixed point values in (\ref{quasi}) translate into the fixed point 
values for the  top-, bottom-quark Yukawa couplings, and the baryon
number violating coupling:
\begin{eqnarray}
\lambda''^{QFP}_{233} & \simeq & 0.90, \nonumber \\
h^{QFP}_b & \simeq & 0.86, \nonumber \\
h^{QFP}_t & \simeq & 0.86.  \label{yquasi}
\end{eqnarray}
These quasi-fixed-point values for the Yukawa couplings are not significantly
different from those obtained in a situation when the $\tau$
Yukawa coupling was ignored, and when the baryon and lepton number
violating couplings 
were considered separately in the fixed point analysis~\cite{bbpw}.
Since the quasi-fixed points are reached for large initial values 
of the couplings at the GUT scale, these reflect on the assumption
of purturbative unitarity, or the  absence of Landau poles, of the
corresponding couplings. The quasi-fixed points (\ref{yquasi}), therefore,
provide an upper bound on the relevant Yukawa and the
baryon and lepton number violating couplings. 
From our analysis we, thus, conclude
that $\lambda''_{233}  \stackrel{<}{\sim}   1$ in a 
process independent manner.

We have carried an analogous numerical study
for the soft supersymmetry breaking trilinear couplings.
Here the choice
of inputs is larger and we economize on the possibilities, however,
ensuring that the main focus of the analysis is not lost.  
In Figs. 2(a), 2(b) and 2(c)
we illustrate the infrared attraction for the trilinear couplings
with a choice of Yukawa and the baryon number violating couplings
$h_b=\lambda''_{233}=1.1$ and
$h_t=1.65$ (in order to break the near isospin invariance of
the top-  and bttom-quark couplings),  and universal boundary conditions
for the trilinear couplings  
$\pm(3,2,1,0.5) M_{1/2},\, 0$, with $M_{1/2}$ the universal gaugino
mass,  for the eight different computations
that we perform.  The results for the Yukawa couplings and the trilinear
couplings at the infrared  scale
are then fed back to the RG equations at the UV scale and then evolved again.  
We exhibit the resulting attraction in terms of the 
projections onto the $\tilde{A}_t - \tilde{A}_b,$ 
$\tilde{A}_t - \tilde{A}_{\lambda''}$ and
$\tilde{A}_b - \tilde{A}_{\lambda''}$ planes in Fig. 2.
The dramatic focussing property of the renormalization group equations
for the $A$ parameters is clearly observed in these results.
The results of this iteration process
results in the following
predictions for the quasi-fixed  points for the $A$ parameters:
\begin{eqnarray}
A_{\lambda''}^{QFP} & \simeq &  0.95 \, m_{\tilde g}, \nonumber\\
A^{QFP}_{t} & \simeq &  0.66 \, m_{\tilde g},  \nonumber\\
A^{QFP}_{b} & \simeq & 0.67 \, m_{\tilde g}, \label{aquasi} 
\end{eqnarray}
where $m_{\tilde g}$ is the gluino mass ($= M_3$) at the weak scale.
These quasi-fixed point values for the $A$ parameters must be
compared with the true fixed point values (\ref{afstable}).
We note that the quasi-fixed-point  values (\ref{aquasi}) 
provide a lower bound on the corresponding $A$ parameters, whereas the
true fixed-point values (\ref{afstable}) represent an upper bound on these
parameters. From this  analysis we are able to constrain these
$A$ parameters in the following model independent
manner:
\begin{eqnarray}
& \displaystyle {{A_{\lambda''}}\over{m_{\tilde g} }}  \simeq  1, 
& \nonumber\\
& \displaystyle 0.66 \stackrel{<}{\sim} {{A_{t}}
\over { m_{\tilde g}}} \stackrel{<}{\sim}  1, 
& \nonumber\\
& \displaystyle 0.67 \stackrel{<}{\sim} {{A_{b}}\over { m_{\tilde g}}}   
\stackrel{<}{\sim} 1,
& \label{abounds}
\end{eqnarray}

\section{SUMMARY AND  CONCLUSIONS}

We have carried out a detailed study of the infrared fixed point structure
of the minimal supersymmetric standard model with the third generation 
Yukawa couplings and with highest generation baryon and lepton 
number violation. We have obtained the infrared fixed points for 
such a model, and shown that there is no physically acceptable
infrared fixed point with both baryon and lepton number violating 
couplings approaching a nontrivial fixed point.  The simultaneous
nontrivial fixed point for the top- and bottom-quark Yukawa couplings,
and the $B$-violating coupling $\lambda''_{233}$ is the only true
fixed point that is stable in the infrared region. 
We have obtained approximate analytical solutions  to the RG equations
near the stable fixed  point, which illustrate the approach to
the fixed point.
We have also derived the exact solutions of the RG equations of such a 
model in a closed form, from which we have obtained infrared 
quasi-fixed point solutions for the various couplings. 
The quasi-fixed-points are realised at the weak scale when the initial 
couplings at the GUT scale are large. These fixed points, thus, reflect
on the assumption of the perturbative unitarity of the 
corresponding couplings.
We have carried out the corresponding RG
analysis of the trilinear soft supersymmetry breaking couplings, and
obtained the infrared stable fixed point for these couplings, and the
approach to the fixed point analytically.

Since the true fixed points may not be reached at the electroweak scale,
we have  also studied the numerical solutions of the RG equations, 
and obtained the infrared fixed surfaces, and demonstrated
the convergence of the
RG flow towards the fixed point. From this analysis we have obtained 
a process independent upper bound on the highest generation
baryon number violating coupling, $\lambda''_{233}  \stackrel{<}{\sim} 1$.

Our study of the RG flow of the Yukawa couplings has been complemented
by the corresponding numerical study of the RG equations for
the soft supersymmetry breaking trilinear couplings, and the demonstration
of the rapid convergence towards the fixed point for these
couplings. From this analysis,
we have constrained the $A$ parameters to be
$A_{\lambda''}/m_{\tilde g}  \simeq  1, \, \,
0.66 \stackrel{<}{\sim} A_{t}/ m_{\tilde g} \stackrel{<}{\sim}  1, \, \,
0.67 \stackrel{<}{\sim} A_{b}/ m_{\tilde g} \stackrel{<}{\sim} 1.$
We emphasize that our results are independent of whether or not there
is a unification of the couplings at some large scale or not.
The infrared fixed points and the infrared fixed surfaces
are the property of the MSSM with baryon and lepton
number violation. Since they are approximate, only their actual value
depends on the large scale.
\bigskip

\noindent{\bf Acknowledgements:}  We thank the organizers of the
WHEPP6, the Sixth Workshop on High Energy Particle Physics, held in
Chennai, India for their hospitality during which this work was started.
BA thanks the Institut f\"ur Theoretische Physik, Universit\"at Bern,
Switzerland for its hospitality when part of this work was done.
The work of PNP is supported by  the University Grants
Commission Research Award No. F.30-63/98 (SA III).  
He would like to thank the Inter-University
Centre for Astronomy and Astrophysics, Pune, India for its hospitality
while part of this work was done.

\newpage

\noindent{\bf Figure Captions}

\bigskip

\noindent {\bf Fig. 1 (a)}  The values of the ratios $R_b$ and
$R_t$ at the end of first four iterations for the inputs 
$h_t=9,1,0.5, \, h_b=9,1,0.5, $ and $\lambda''_{233}=1.1$.

\medskip

\noindent {\bf Fig. 1 (b)}  The values of the ratios $R''$ and
$R_t$ at the end of first four iterations for the inputs 
$h_t=9,1,0.5, \, h_b=9,1,0.5, $ and $\lambda''_{233}=1.1$.

\medskip

\noindent {\bf Fig. 1 (c)}  The values of the ratios $R''$ and
$R_b$ at the end of first four iterations for the inputs 
$h_t=9,1,0.5, \, h_b=9,1,0.5, $ and $\lambda''_{233}=1.1$.

\medskip

\noindent {\bf Fig. 2 (a)}  The values of the ratios $\tilde{A}_b$ and
$\tilde{A}_t$ at the end of first four iterations for the inputs 
$h_b=\lambda''_{233}=1.1$ and $h_t=1.65$, and universal trilinear couplings
$A=\pm(3,2,1,0.5) M_{1/2}, \, 0$.

\medskip

\noindent {\bf Fig. 2 (b)}  The values of the ratios $\tilde{A}_{\lambda''}$ and
$\tilde{A}_t$ at the end of first four iterations for the inputs 
$h_b=\lambda''_{233}=1.1$ and $h_t=1.65$, and universal trilinear couplings
$A=\pm(3,2,1,0.5) M_{1/2}, \, 0$.

\medskip

\noindent {\bf Fig. 2 (c)}  The values of the ratios $\tilde{A}_{\lambda''}$ and
$\tilde{A}_b$ at the end of first four iterations for the inputs 
$h_b=\lambda''_{233}=1.1$ and $h_t=1.65$, and universal trilinear couplings
$A=\pm(3,2,1,0.5) M_{1/2}, \, 0$.

\newpage

\begin{center}
\begin{figure}\nonumber
\epsfig{figure=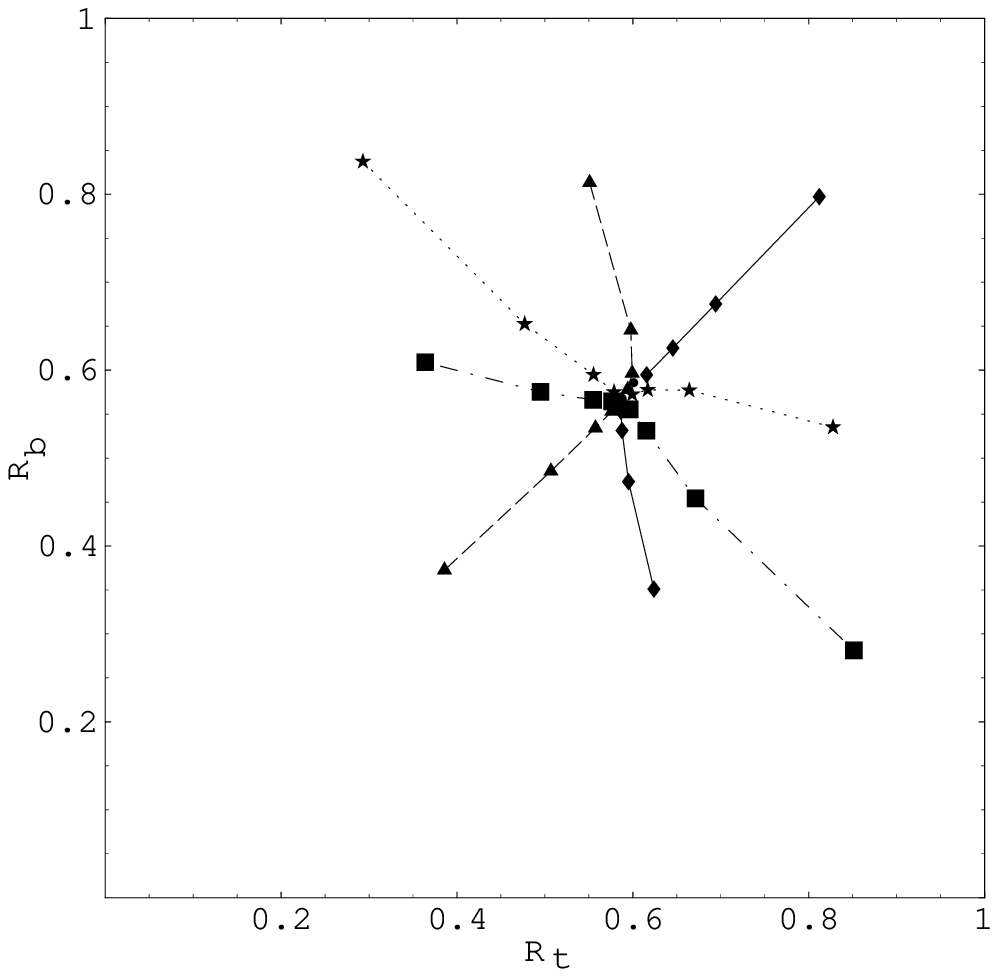,width=10cm,height=10cm}
\centerline{Fig. 1 (a)}
\end{figure}
\begin{figure}\nonumber
\epsfig{figure=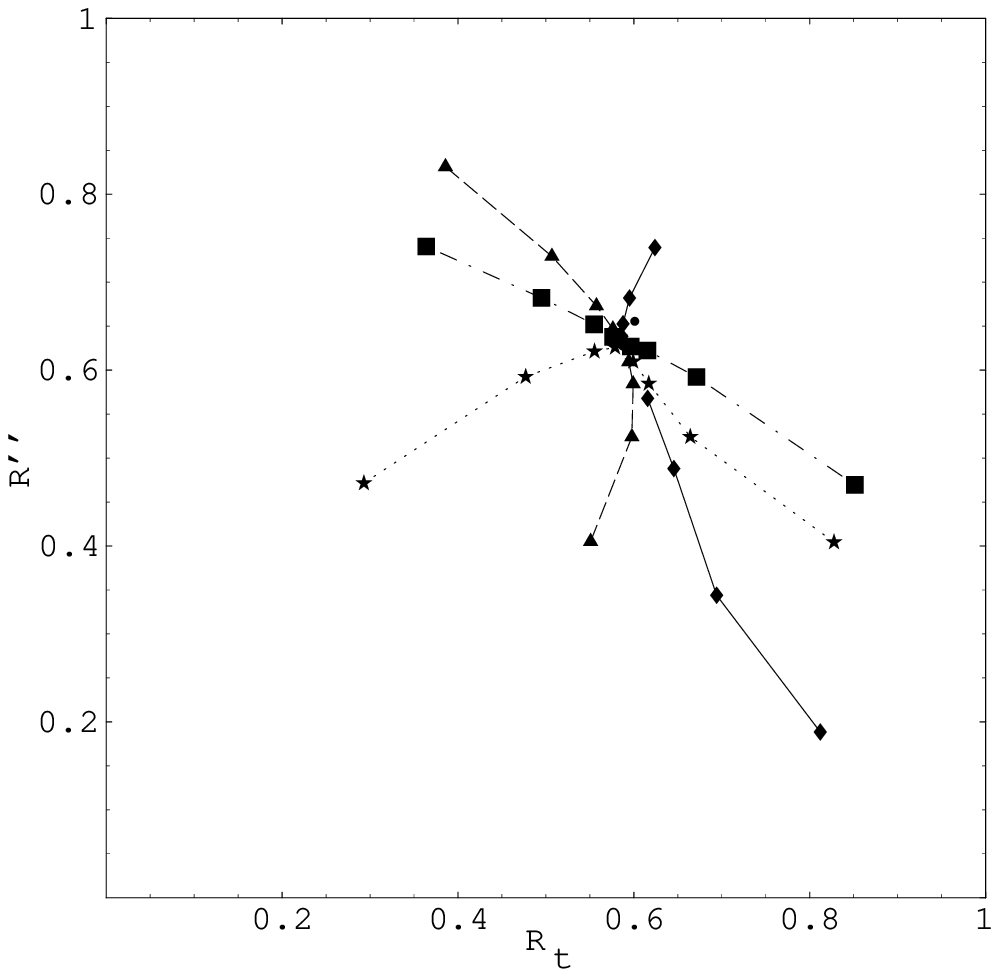,width=10cm,height=10cm}
\centerline{Fig. 1 (b)}
\end{figure}

\begin{figure}\nonumber
\epsfig{figure=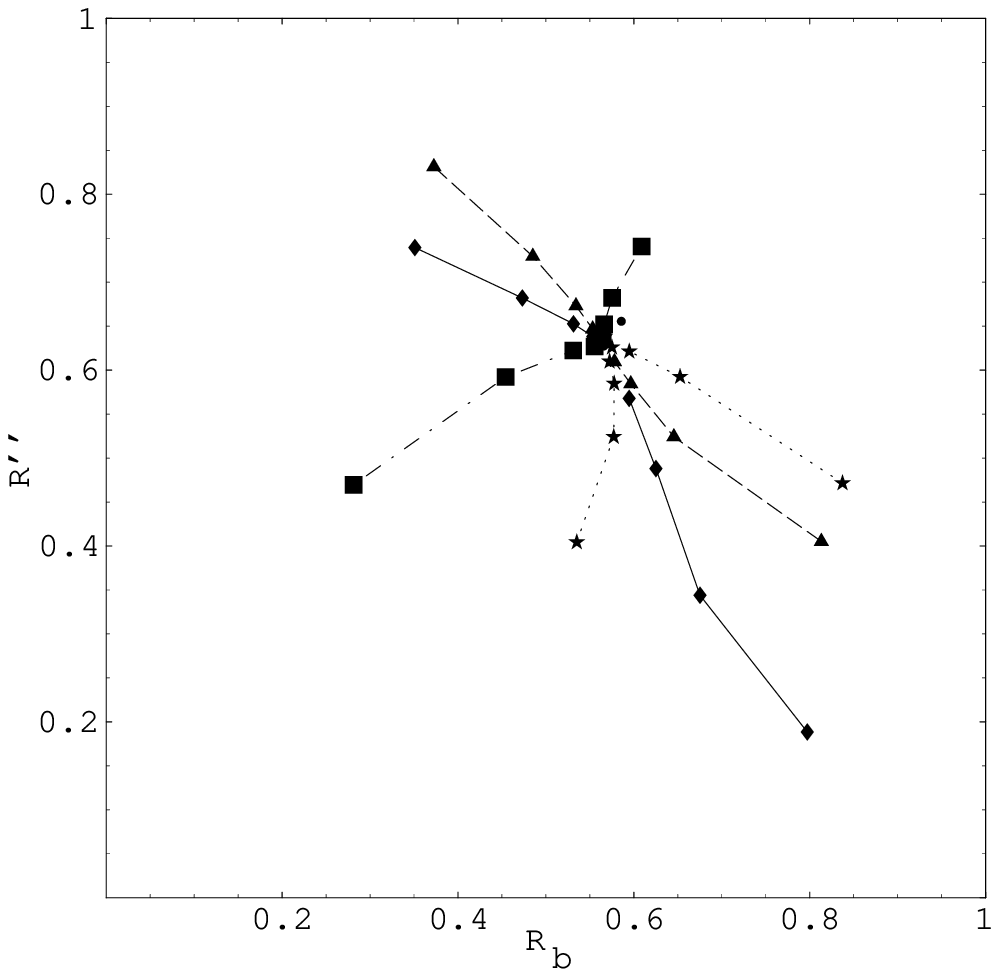,width=10cm,height=10cm}
\centerline{Fig. 1 (c)}
\end{figure}

\newpage

\begin{figure}\nonumber
\epsfig{figure=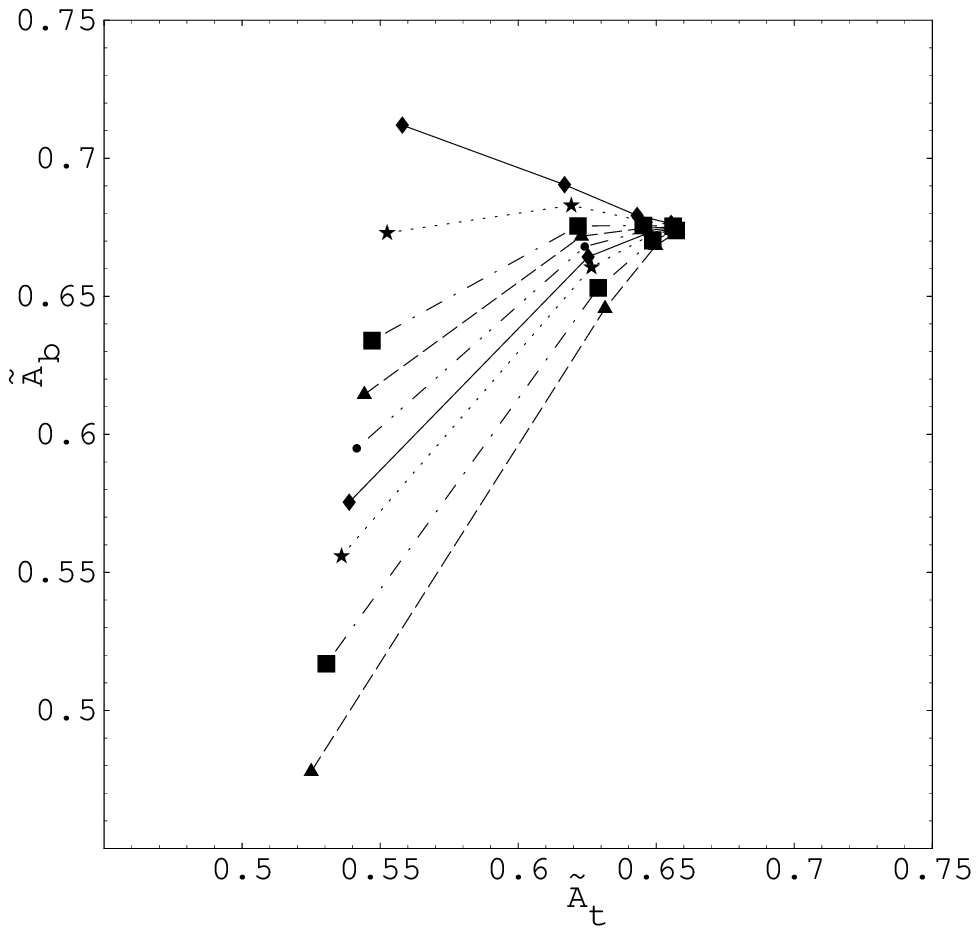,width=10cm,height=10cm}
\centerline{Fig. 2 (a)}
\end{figure}
\begin{figure}\nonumber
\epsfig{figure=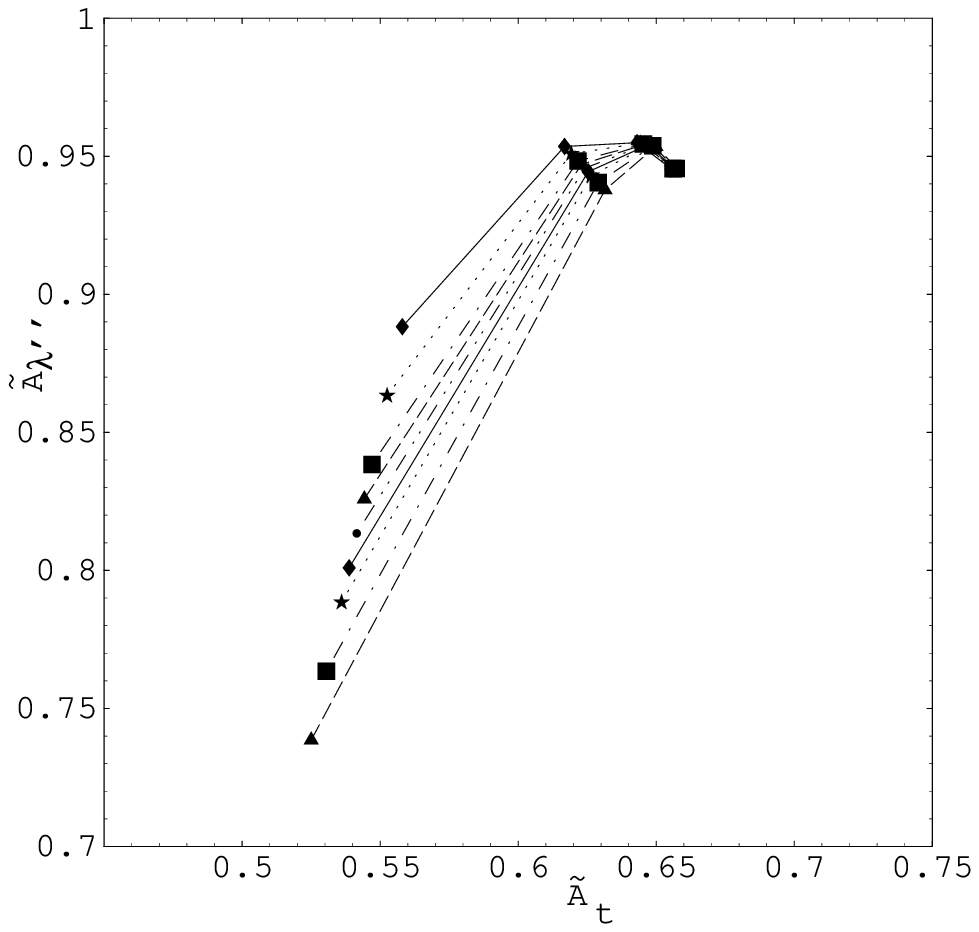,width=10cm,height=10cm}
\centerline{Fig. 2 (b)}

\end{figure}
\begin{figure}\nonumber
\epsfig{figure=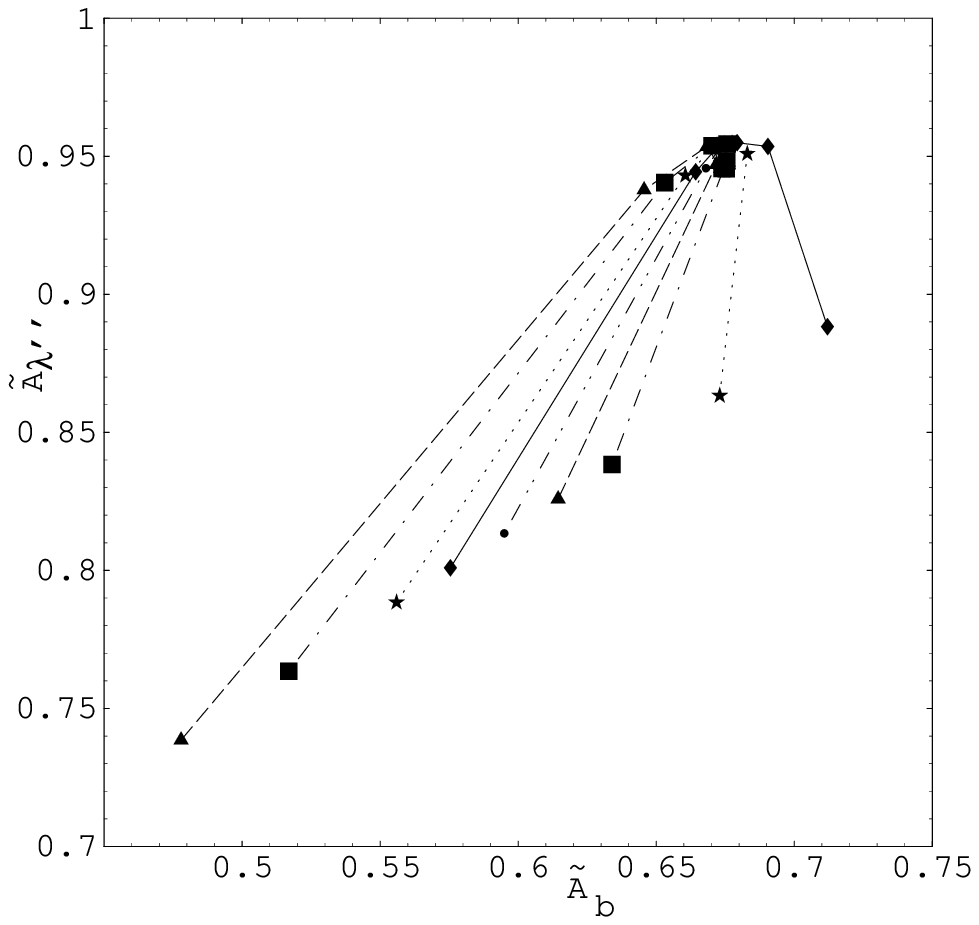,width=10cm,height=10cm}
\centerline{Fig. 2 (c)}
\end{figure}

\end{center}


\begin{thebibliography}{abcde}

\bibitem{bsmw1}
B.~Schrempp and M.~Wimmer,
Prog.\ Part.\ Nucl.\ Phys.\  {\bf 37}, 1 (1996).


\bibitem{bpggr1}
B.~Pendleton and G.~G.~Ross,
Phys.\ Lett.\  {\bf B98}, 291 (1981).

\bibitem{mlggr1}
M.~Lanzagorta and G.~G.~Ross,
Phys.\ Lett.\  {\bf B349}, 319 (1995).



\bibitem{cth1}
C.~T.~Hill,
Phys.\ Rev.\  {\bf D24}, 691 (1981).

\bibitem{wsy}
S.~Weinberg,
Phys.\ Rev.\  {\bf D26}, 287 (1982);
N.~Sakai and T.~Yanagida,
Nucl.\ Phys.\  {\bf B197}, 533 (1982).


\bibitem{ff}
G.~R.~Farrar and P.~Fayet,
Phys.\ Lett.\  {\bf B76}, 575 (1978).


\bibitem{bbpw}
V.~Barger, M.~S.~Berger, R.~J.~Phillips and T.~Wohrmann,
Phys.\ Rev.\  {\bf D53}, 6407 (1996);\\
see also
B.~Brahmachari and P.~Roy,
Phys.\ Rev.\  {\bf D50}, 39 (1994).

\bibitem{bapnp1}
B.~Ananthanarayan and P.~N.~Pandita,
Phys.\ Lett.\  {\bf B454}, 84 (1999).

\bibitem{bapnp2}
B.~Ananthanarayan and P.~N.~Pandita,
Phys.\ Rev.\  {\bf D62}, 036009 (2000) and references therein.

\bibitem{add1}
B.~C.~Allanach, A.~Dedes and H.~K.~Dreiner,
Phys.\ Rev.\  {\bf D60}, 056002 (1999).



\bibitem{barbier}
R.~Barbier {\it et al.},
hep-ph/9810232.

\bibitem{auberson1}
G.~Auberson and G.~Moultaka,
Eur.\ Phys.\ J.\  {\bf C12}, 331 (2000).

\bibitem{bs1}
B.~Schrempp,
Phys.\ Lett.\  {\bf B344}, 193 (1995).

\bibitem{groom}
D.~E.~Groom {\it et al.},
Eur.\ Phys.\ J.\  {\bf C15}, 1 (2000).



\bibitem{kazakov1}
D.~Kazakov and G.~Moultaka,
Nucl.\ Phys.\  {\bf B577}, 121 (2000).


\bibitem{ijdrtj1}
I.~Jack and D.~R.~T.~Jones,
Phys.\ Lett.\  {\bf B443}, 177 (1998).



\bibitem{codkaz1}
S.~Codoban and D.~I.~Kazakov,
Eur.\ Phys.\ J.\  {\bf C13}, 671 (2000).

\end{thebibliography}
\end{document}